\documentclass[12pt,prd]{article}
\pdfoutput=1
\usepackage{jheppub}
\usepackage{rotating}
\usepackage{array}
\usepackage{amsmath}
\usepackage[normalem]{ulem}
\usepackage{slashed}
\usepackage{booktabs}
\usepackage[pdftex,table]{xcolor}
\usepackage{units}
\usepackage{xfrac}
\usepackage{mathtools}
\usepackage{empheq}
\usepackage[]{units}
\usepackage{multirow}
\usepackage{amssymb}
\usepackage{url}
\usepackage{comment}
\usepackage{paralist}

\usepackage{tikz}

\def\beq{\begin{equation}}
\def\eeq{\end{equation}}
\newcommand{\bea}{\begin{eqnarray}\begin{aligned}}
\newcommand{\eea}{\end{aligned}\end{eqnarray}}
\def\bitem{\begin{itemize}}
\def\eitem{\end{itemize}}

\definecolor{darkpurple}{rgb}{0.5, 0.2, 0.8}
\definecolor{darkblue}{rgb}{0.0, 0.0, 0.8}
\definecolor{darkgreen}{rgb}{0.0, 0.4, 0.0}
\definecolor{darkred}{rgb}{0.5, 0.0, 0.0}
\usepackage{hyperref}
\hypersetup{
    linktocpage,
     colorlinks,
     citecolor=darkgreen,
     linkcolor= darkgreen,
     urlcolor=darkgreen
}
\usepackage{lineno}

\abstract{
Jet clustering is traditionally an unsupervised learning task because there is no unique way to associate hadronic final states with the quark and gluon degrees of freedom that generated them.  However, for uncolored particles like $W$, $Z$, and Higgs bosons, it is possible to approximately (though not exactly) associate final state hadrons to their ancestor.  By labeling simulated final state hadrons as descending from an uncolored particle, it is possible to train a supervised learning method to create boson jets.  Such a method would operate on individual particles and identify connections between particles originating from the same uncolored particle.  Graph neural networks are well-suited for this purpose as they can act on unordered sets and naturally create strong connections between particles with the same label.  These networks are used to train a supervised jet clustering algorithm.  The kinematic properties of these graph jets better match the properties of simulated Lorentz-boosted $W$  bosons.  Furthermore, the graph jets contain more information for discriminating $W$ jets from generic quark jets.  This work marks the beginning of a new exploration in jet physics to use machine learning to optimize the construction of jets and not only the observables computed from jet constituents.
}


\begin{document}
\title{Supervised Jet Clustering with Graph Neural Networks for Lorentz Boosted Bosons}

\author{Xiangyang Ju} 
\author{and Benjamin Nachman}
\affiliation{\normalsize Physics Division, Lawrence Berkeley National Laboratory, Berkeley, CA 94720, USA}

\emailAdd{xju@lbl.gov}
\emailAdd{bpnachman@lbl.gov}

\maketitle
 
\section{Introduction}\label{sec:intro}

Lorentz-boosted massive bosons are a common feature of theories that extend the Standard Model (SM) of particle physics.  In particular, new heavy particles introduced to solve one of the challenges with the SM may predominately decay into bosons and if there is a large mass hierarchy  between the heavy particle and the bosons, the latter will be produced in the lab frame with a significant Lorentz boost.  Singly produced bosons can also have significant Lorentz boost when produced in association with initial state radiation.  The ATLAS and CMS collaborations have performed extensive searches involving boosted bosons decaying hadronically in the $VV$~\cite{Aad:2019fbh,Aad:2020ddw,Sirunyan:2019jbg,Sirunyan:2019vgt}, $Vh$~\cite{Aad:2020tps,Sirunyan:2018fuh}, $hh$~\cite{Aad:2020ldt,Sirunyan:2019quj,Sirunyan:2017isc}, $VX$~\cite{Aad:2020hzm},
$Xh$~\cite{Aaboud:2017ecz},
$XY$~\cite{Aad:2020cws}, single-$X$~\cite{Aaboud:2018zba,Sirunyan:2017dnz,Sirunyan:2018ikr,Sirunyan:2019sgo,Sirunyan:2019vxa}, and single-$h$~\cite{ATLAS-CONF-2018-052,Sirunyan:2017dgc} channels, where $V\in\{W^\pm,Z\}$, $h$ is the SM Higgs boson, and $X/Y$ are beyond the SM bosons.

A variety of jet substructure techniques have been developed to enhance Lorentz boosted boson tagging~\cite{1803.06991,1709.04464,Adams:2015hiv,Altheimer:2013yza,Abdesselam:2010pt,Altheimer:2012mn,Salam:2009jx,Marzani:2019hun}.  These methods range from physically motivated features such as groomed jet mass~\cite{Butterworth:2008iy,Ellis:2009me,Krohn:2009th,Larkoski:2014wba,Dasgupta:2013ihk}, $N$-subjettiness~\cite{Thaler:2010tr,Thaler:2011gf} and $D_2$~\cite{Larkoski:2014gra} to complex observables built using machine learning~\cite{1709.04464}.  ATLAS and CMS have integrated and extended these methods as well as studied them using collision data~\cite{Aad:2019uoz,Aaboud:2018psm,Aaboud:2018kfi,Sirunyan:2020lcu,CMS-PAS-JME-16-003}.  One feature that all of these algorithms have in common is that they start from a collection of constituents selected using a jet clustering algorithm.  Various studies have investigated optimizing the jet clustering algorithm by considering many options~\cite{ATL-PHYS-PUB-2019-027,ATLAS-CONF-2020-021,Chakraborty:2020vwj}.  While important for converging on a method in the traditional paradigm, these approaches are fundamentally limited by the discreteness of the algorithm types and the flexibility offered by the tunable parameters of a given algorithm.  

The most common approach for forming the initial Lorentz boosted boson candidate jets is the anti-$k_t$ algorithm~\cite{Cacciari:2008gp}.  This algorithm is a form of \textit{unsupervised} learning because no per-particle labels are used to form the jets\footnote{Previous attempts at combining jet finding with \textit{unsupervised} machine learning have been studied in the past~\cite{Mackey:2015hwa,Grigoriev:2003tn}, but do not have the benefits of the supervised approaches discussed here.}.  Instead, a distance measure motivated by the fragmentation of quarks and gluons is used to collect constituents that were likely produced from the same initiating high-energy quark or gluon.  This last sentence does not have a precise meaning because quark and gluon jets are not well-defined objects~\cite{Gras:2017jty,1810.05653}.  Due to the strength of the strong force, the energy flows from outgoing quarks and gluons are interconnected with each other and with the beam remnants.  In contrast, the quarks and gluons from color singlet massive bosons are isolated from the rest of the event.  In the limit that the number of colors $N_c\rightarrow\infty$ or the width of the boson resonance $\Gamma\rightarrow 0$, there is a unique mapping between final state hadrons and ancestor color singlet.  The corrections to this picture are suppressed by at least $(1/N_c)^2$ (`color reconnection') and by powers of $\Gamma/\Lambda_\text{QCD}$.  

Given the approximate (but not exact) mapping between hadrons and color singlets, it makes sense to ask if one could construct a \textit{supervised} approach to forming jets.  In particular, a machine could be trained to label individual particles as originating from a color singlet or not based on the particle kinematic properties as well as the relationship with other particles in the event.  While such an approach may give up the calculability afforded by algorithms like anti-$k_t$, it may provide an optimal approach to constructing jets for searches where calculablility is not necessarily required.  If the jets are constructed optimally, then their substructure should contain as much information as possible for identifying their origin.  One could even co-optimize the jet construction and the jet classification in an end-to-end approach~\cite{Andrews:2019faz,Andrews:2018nwy}, but there are many benefits to first building jets, such as the jet energy calibration. 

Modern machine learning has proven to be a powerful toolkit for jet substructure. For example, a wide range of architectures and applications have been studied for tagging the origin of jets~\cite{Larkoski:2017jix,Guest:2018yhq,Kasieczka:2019dbj,hepmllivingreview,Cogan:2014oua,Almeida:2015jua,deOliveira:2015xxd,ATL-PHYS-PUB-2017-017,Lin:2018cin,Komiske:2018oaa,Barnard:2016qma,Komiske:2016rsd,Kasieczka:2017nvn,Macaluso:2018tck,Nguyen:2018ugw,ATL-PHYS-PUB-2019-028,Andrews:2018nwy,Guest:2016iqz,Louppe:2017ipp,Cheng:2017rdo,Henrion:DLPS2017,Ju:2020xty,Martinez:2018fwc,Moreno:2019bmu,Qasim:2019otl,Chakraborty:2019imr,Chakraborty:2020yfc,1801423,Komiske:2018cqr,Qu:2019gqs,Datta:2019,Datta:2017rhs,Datta:2017lxt,Komiske:2017aww,Butter:2017cot,Chen:2019uar,Fraser:2018ieu,Datta:2019ndh,Moreno:2019neq,Stoye:DLPS2017,Chien:2018dfn,Kasieczka:2018lwf,1806025,Diefenbacher:2019ezd,Nakai:2020kuu,Sirunyan:2017ezt,bielkov2020identifying,Baldi:2014kfa,10.1088/2632-2153/ab9023,1792136,deOliveira:2018lqd,Paganini:DLPS2017,Hooberman:DLPS2017,Belayneh:2019vyx}.  To construct a supervised jet clustering algorithm, a machine learning architecture is needed that can process variable length sets as input.  Multiple such \textit{point cloud} methods have been studied for jet substructure~\cite{Komiske:2018cqr,Qu:2019gqs,Bernreuther:2020vhm,Martinez:2018fwc,Moreno:2019bmu,Henrion:DLPS2017}, but the structure chosen here is the graph neural network (GNN) (see Ref.~\cite{Qu:2019gqs,Ju:2020xty,Bernreuther:2020vhm,Martinez:2018fwc,Moreno:2019bmu,Henrion:DLPS2017,1797439,Ren:2019xhp,Abdughani:2018wrw,Shlomi:2020gdn,Choma:2020cry}).  This is because GNNs not only can process variable length sets, but they can also label the relationship between elements (not unique to GNNs, but natural given their construction).  This property is critical for labeling particles as originating from the color singlet ancestor or not. Labeling constituents is also known as \textit{semantic segmentation} and has been studied for other tasks in high energy physics ranging from pileup particle identification~\cite{Martinez:2018fwc,Komiske:2017ubm} to liquid argon time projection chamber labeling~\cite{Adams:2018bvi,Koh:2020snv}. In addition, a recent study~\cite{Iiyama:2020wap} shows that GNNs can be executed with a latency of less than 1~$\mu$s on an field-programmable gate arrays, making such networks very promising for real-time data learning and filtering.

This paper is organized as follows.  Section~\ref{sec:sim} introduces the simulated samples used to train the supervised jet clustering algorithm, where Lorentz-boosted $W$ bosons provide a reoccurring example.  The graph neural network methods are described in Section~\ref{sec:ML} and numerical results are presented in Section~\ref{sec:results}.  The paper ends with outlook and conclusions in Section~\ref{sec:conclusions}.

\section{Simulation}\label{sec:sim}

Proton-proton collisions are simulated with \textsc{Pythia} 8.183~\cite{Sjostrand:2007gs,Sjostrand:2006za} at a center-of-mass-energy of $\sqrt{s}=13$ TeV.  Lorentz boosted $W$ bosons are generated from the decay of a hypothetical $W'$ boson with a mass of 600 GeV that decays 100\% of the time to a $W$ boson and a $Z$ boson.  The $W$ boson is forced to decay hadronically and the $Z$ boson decays into neutrinos.  To simulate a quark jet with nearly the same kinematic properties, a hypothetical excited quark $q^*$ with a mass of 600 GeV is generated and decays 100\% of the time into a quark and a $Z$ boson.  This $Z$ boson then is forced to decay into neutrinos.  The widths of the $W', q^*$, and $W$ boson are set to 0.01 GeV.  In total, 100,000 $W'$ and $q^*$ events were generated.

\begin{figure}[h!]
    \centering
    \includegraphics[width=0.95\textwidth]{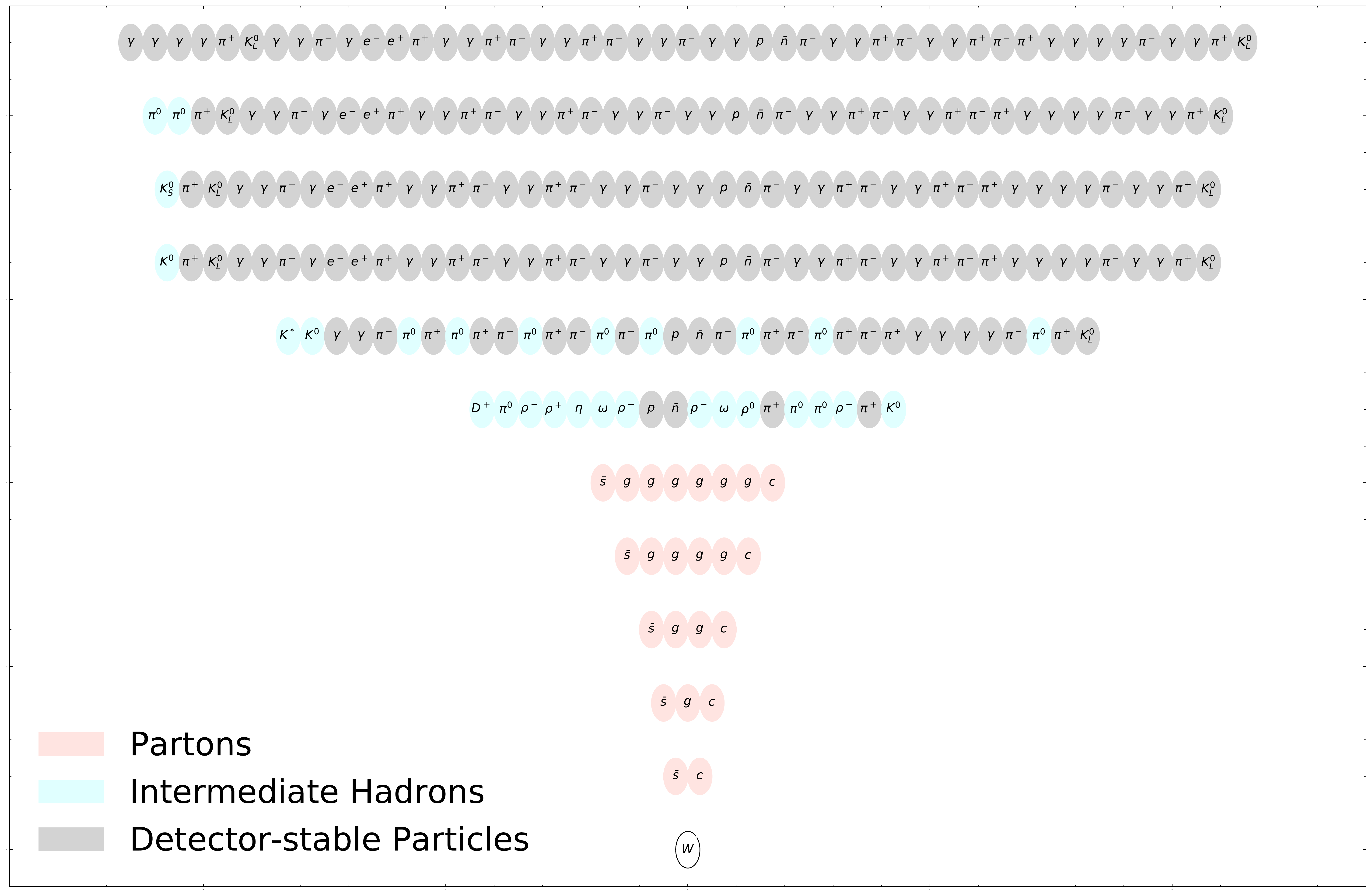}
    \caption{An illustration of the $W\rightarrow c\bar{s}$ decay tracing for a single event.  At each step, every non-detector-stable particle is replaced with their immediate descendants from the Pythia event record.  The order per row is arbitrary.}
    \label{fig:decay}
\end{figure}

As a leading $N_c$ generator such as \textsc{Pythia}, it is possible to uniquely trace final state hadrons to the $W$ boson.  Individual final state hadrons are then labeled based on the existence (or not) of a real $W$ boson in their ancestry from the event record.   This is illustrated for one event in Fig.~\ref{fig:decay}.

To compare with the graph neural network-based clustering scheme described in the next section, jets are clustered using the anti-$k_t$ algorithm~\cite{Cacciari:2008gp} with radius parameter $R = 1.0$ implemented in \textsc{Fastjet} 3.0.3~\cite{Cacciari:2011ma,Cacciari:2005hq}.  Jets are only kept if they have $p_T>100$ GeV.  These jets are subsequently trimmed~\cite{Krohn:2009th} by keeping only $R=0.2$ subjets with at least $5\%$ of the ungroomed jet's transverse momentum.  Trimming is not the only jet grooming algorithm~\cite{Butterworth:2008iy,Ellis:2009me,Krohn:2009th,Larkoski:2014wba,Dasgupta:2013ihk}, but it is widely used (see e.g. Ref.~\cite{ATL-PHYS-PUB-2019-027,ATLAS-CONF-2020-021}).

\begin{figure}[h!]
    \centering
    \includegraphics[height=0.45\textwidth]{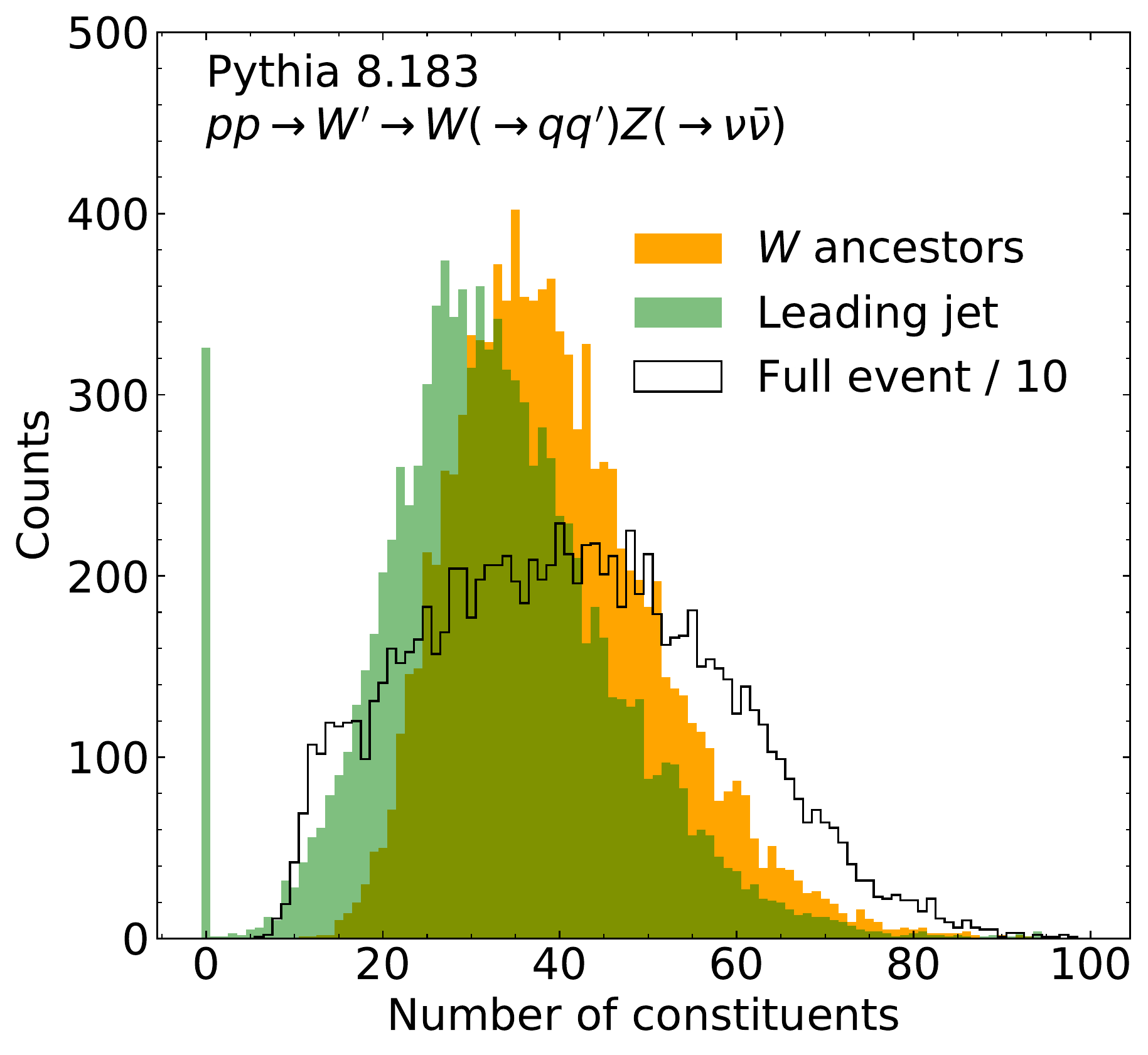}
    \includegraphics[height=0.45\textwidth]{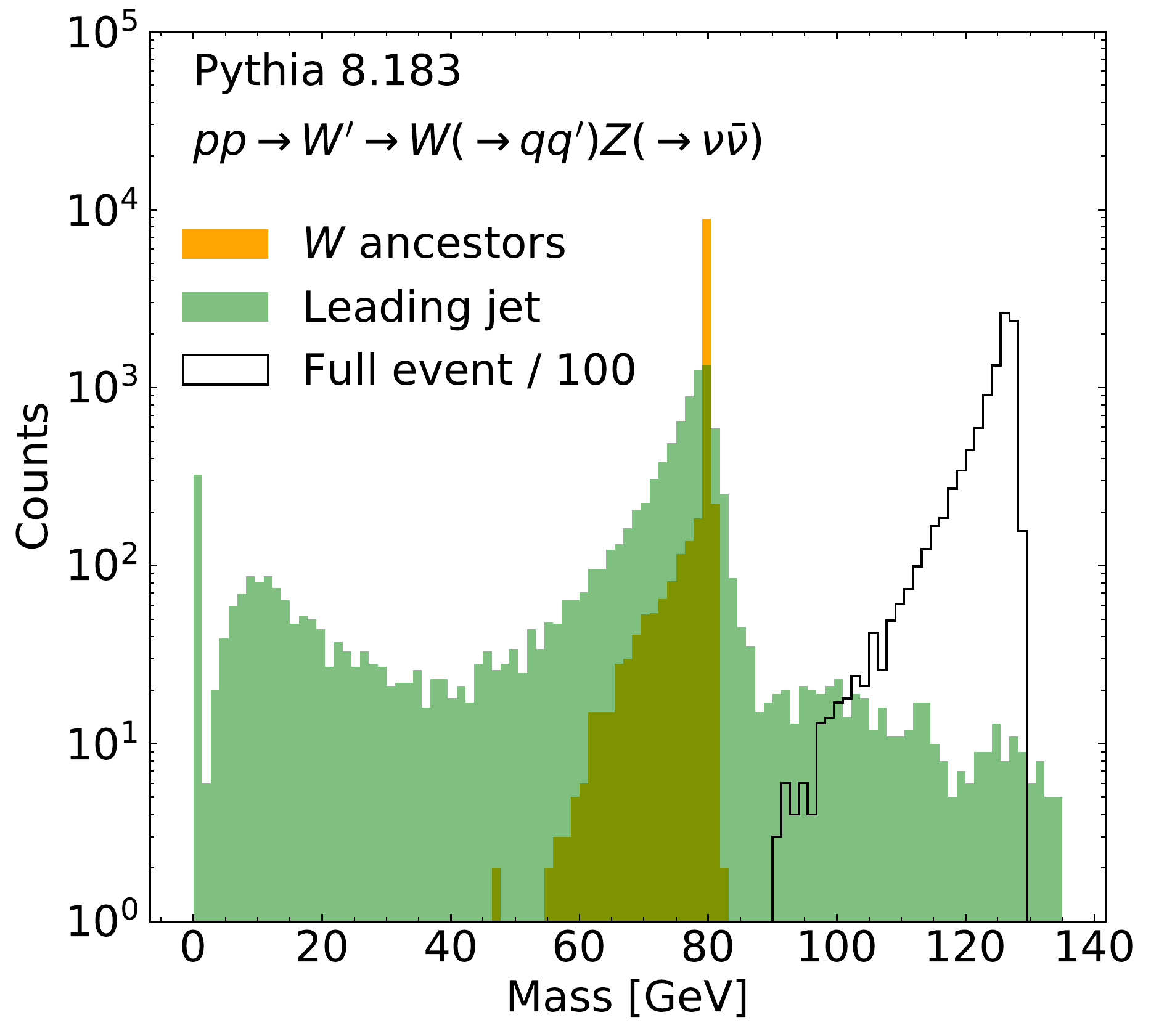}
    \caption{Left: a histogram of the number of detector-stable particles originating from the $W$ boson, inside the leading jet, and in the  full event from the $W'$ production. The leading jet is constructed from the anti-$k_t$ algorithm. The spike at 0 corresponds to events with no jet with $p_T>100$ GeV. For the full event, the number of constituents is divided by 10.  Right: a histogram of the mass from detector-stable particles originating from the $W$ boson, inside the leading jet, and in the full event from the $W'$ production. For the full event, the mass is divided by 100. }
    \label{fig:basicplots}
\end{figure}

\begin{figure}[h!]
    \centering
    \includegraphics[height=0.45\textwidth]{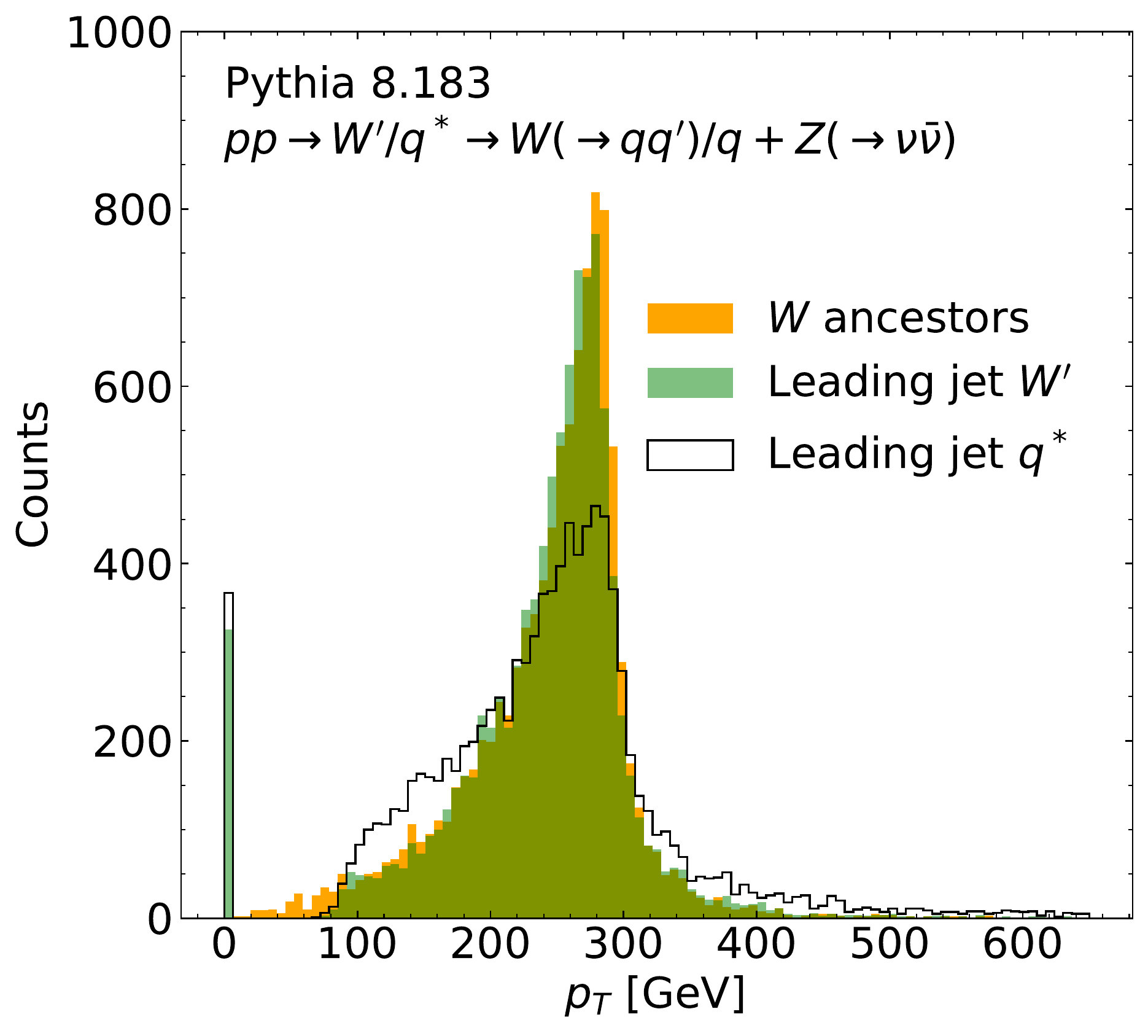}
    \caption{A histogram of the $p_T$ of the vector sum of the four momenta of the detector-stable particles originating from the $W$ boson and inside the leading jet in the $W'$ and $q^*$ events.}
    \label{fig:basicplotspT}
\end{figure}

Figure~\ref{fig:basicplots} presents histograms of basic quantities in $W'$ events.  The number of detector-stable particles with a $W$ ancestor is about the same as the number of constituents inside the leading jet clustered by the anti-$k_t$ algorithm, however, it only accounts for about 10\% of the total number of detector-stable particles in the event.  The mass computed from the detector-stable particles originating from a $W$ boson is nearly exactly $m_W$ while leading jet mass is peaked around $m_W$ with a broad width. On the other hand, there are many non-$W$ particles in the event, giving rise to an event mass far from the $W$ boson mass. Therefore, it is non-trivial  for a machine to find the $W$ decay products in order to reconstruct the $W$ boson mass. In the leading jet case, the low-mass peak corresponds to cases where both quarks from the $W$ decay are not mostly contained within the leading jet or the leading jet is unrelated to the quarks from the $W$ decay.  Figure~\ref{fig:basicplotspT} shows that the kinematic properties of the jets in $W'$ and $q^*$ events are similar.  The jet transverse momentum spectra are not identical because the radiation pattern outside of the jet cone is different for the color singlet $W$ and color triplet quarks. 

\section{Graph Neural Network Methods}\label{sec:ML}
A graph contains a set of nodes, a set of edges with each connecting a pair of nodes, and a set of node-, edge- and graph-level attributes, collectively called graph attributes. Graph Neural Networks (GNN) are trainable functions that operate on a graph to learn latent graph attributes as well as to form a parameterized message-passing by which information is propagated across the graph, ultimately learning sophisticated graph attributes.

Each collision event is represented as a fully connected bidirectional graph in which the nodes are the final state particles and the edges are the connections between all pairs of particles. The node-level attributes are the four-momenta of the particles and the edge- and graph-level attributes will be learned by a GNN. The GNN architecture is same as the one in Ref.~\cite{Ju:2020xty}, which is based on the model in Ref.~\cite{battaglia2018relational}, composed of four trainable components: \begin{inparaenum}[1)]
\item a node encoder which transforms the node-level attributes into their latent representations;
\item an edge encoder which transforms the aggregated latent attributes of its neighbouring nodes into their latent representations;
\item an interaction network~\cite{Battaglia:2016jem};
\item and an decoder that computes graph- or edge-level classification scores.
\end{inparaenum}
The encoders and the decoder use basic deep learning building blocks including multilayer perceptrons.

The boosted $W$ boson is reconstructed by training a GNN, namely the \textit{edge classifier}, to learn the relational information of the final state hadrons. Specifically, the edge-level attributes of the simulated $W$ boson events are labeled as 1 if two hadrons come from the same $W$ boson and 0 otherwise. The edge classifier outputs edge-level classification scores, abbreviated as edge scores, which are compared with the edge labeling using the binary cross-entropy loss. Trainable parameters in the classifier are optimized by the gradient-based stochastic optimizer, Adam~\cite{adam}. The reconstructed $W$ boson candidate for each event is built from the hadrons that are connected by edges with scores larger than a threshold of 0.5. The threshold is a hyper-parameter that can be tuned for a specific problem. The four-momenta of the reconstructed $W$ boson candidate are the sum of the four-momenta of the selected hadrons. The ``edge classifier'' was trained with 90,000 simulated $W$ boson events and tested with 5,000 events.

The reconstructed $W$ boson candidates from the GNN-based edge classifier carry unique information which other machine learning architectures (or traditional jet substructure observables) can use in order to separate the $W$ boson events from background events, such as the $q^*$ events. In this study, another GNN with the same architecture as the edge classifier is used, namely the \textit{event classifier}. The input graphs are the fully connected bidirectional graphs constructed from the hadrons selected by the trained edge classifier. The graph-level attributes are labeled as 1 for the $W$ boson events and 0 for the $q^*$ events. The event classifier outputs the graph-level classification score, abbreviated as event scores, which are compared with the graph labeling using the binary cross-entropy loss. Trainable parameters in the classifier are optimized by the gradient-based stochastic optimizer, Adam. The event classifier was trained with 90,000 $W$ boson events and 90,000 $q^*$ events, and tested with other 5,000 $W$ boson events and 5,000 $q^*$ events. As a comparison, the GNN is also trained with the inputs from the anti-$k_t$ algorithm. In this case, the input graphs are the fully connected bidirectional graphs constructed from the hadrons inside the leading jet which in turn is constructed from the anti-$k_t$ algorithm. To facilitate the discussions below, the GNN trained with the inputs from the anti-$k_t$ algorithm is called \textit{tGNN} while that trained with the inputs from the trained edge classifier is called \textit{eGNN}.  All training was performed on an NVIDIA V100 GPU.

\section{Results}\label{sec:results}
\begin{figure}[htb]
    \centering
    \includegraphics[width=\linewidth]{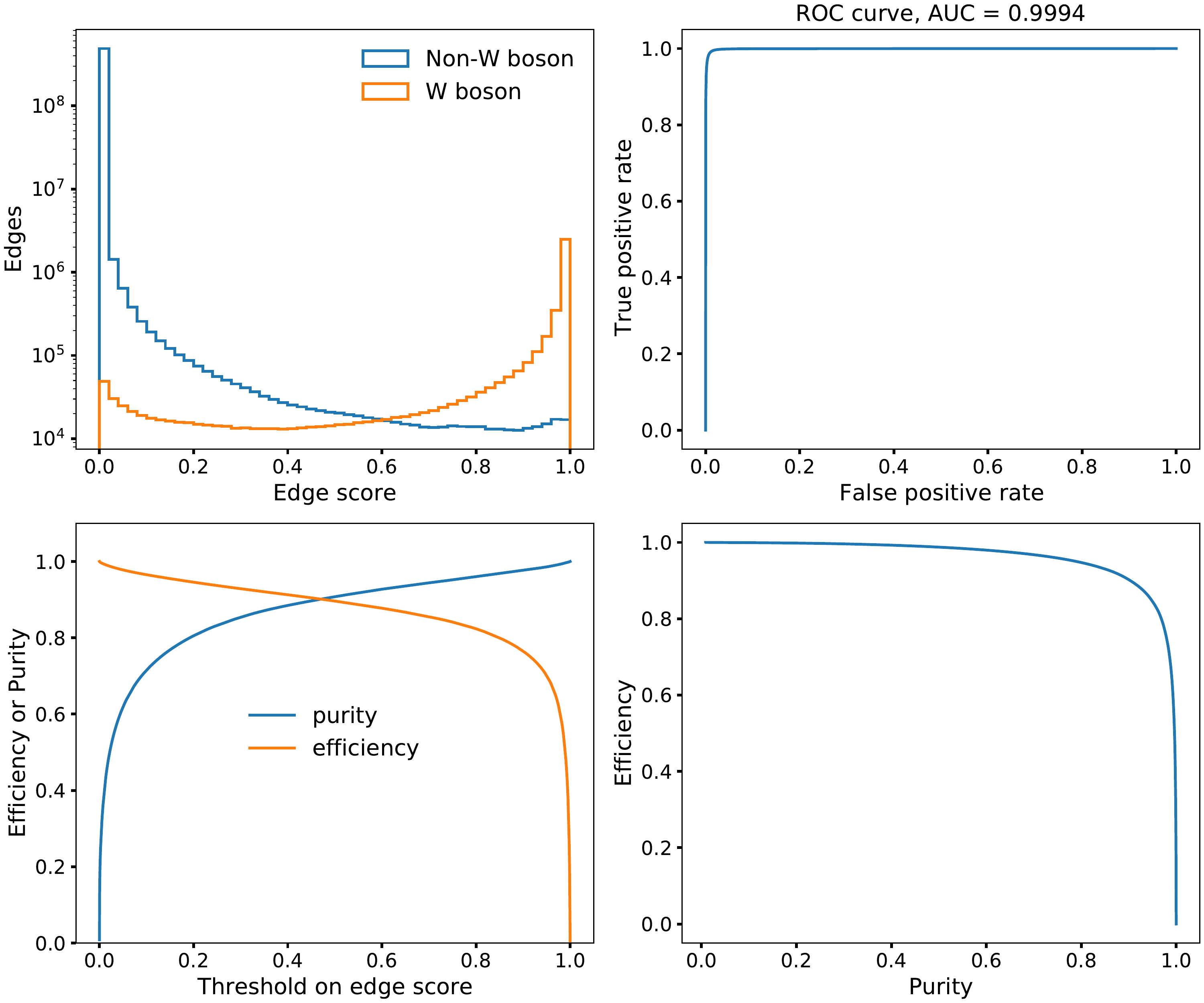}
    \caption{Metrics used in evaluating classification performance of the edge classifier. 
    Upper Left: the distribution of the edge score for edges that connect the hadrons coming from the $W$ boson in yellow (true edges) and the edges that do not in blue (fake edges). 
    Upper Right: the receiver operating characteristic (ROC) curve. AUC is the area under the ROC curve. 
    Bottom Left: the edge efficiency and edge purity as a function of the threshold on the edge score. The definition of the edge efficiency and purity can be found in the text. 
    Bottom Right: the edge efficiency versus the edge purity.     
    }
    \label{fig:edge_scores}
\end{figure}

The edge classifier was trained for 30 epochs, after which no improvement was seen when the model was evaluated on the testing data. The performance of the edge classifier is showed in Figure~\ref{fig:edge_scores}. Two important metrics are the edge efficiency, defined as the ratio of the number of true edges passing the threshold over the number of total true edges, and the purity, defined as the ratio of the number of true edges passing the threshold over the number of total edges passing the threshold.  Varying the threshold in the edge scores results in different values of edge efficiency and purity.  Table~\ref{tab:eff_purity} shows the edge efficiency and purity for three different thresholds on the edges scores.

\begin{table}[htb]
    \caption{The edge efficiency and edge purity as a function of the threshold on the edge scores. The definition of the edge efficiency and edge purity can be found in the text.    \label{tab:eff_purity}}
    \centering
    \begin{tabular}{c|c|c|c}
    \toprule
        Threshold & 0.1 & 0.5 & 0.8 \\
        Edge Efficiency & 0.965 & 0.896 & 0.824 \\
        Edge Purity & 0.715 & 0.908 & 0.960 \\
    \bottomrule
    \end{tabular}
\end{table}

\begin{figure}[htb]
    \centering
    \includegraphics[width=\linewidth]{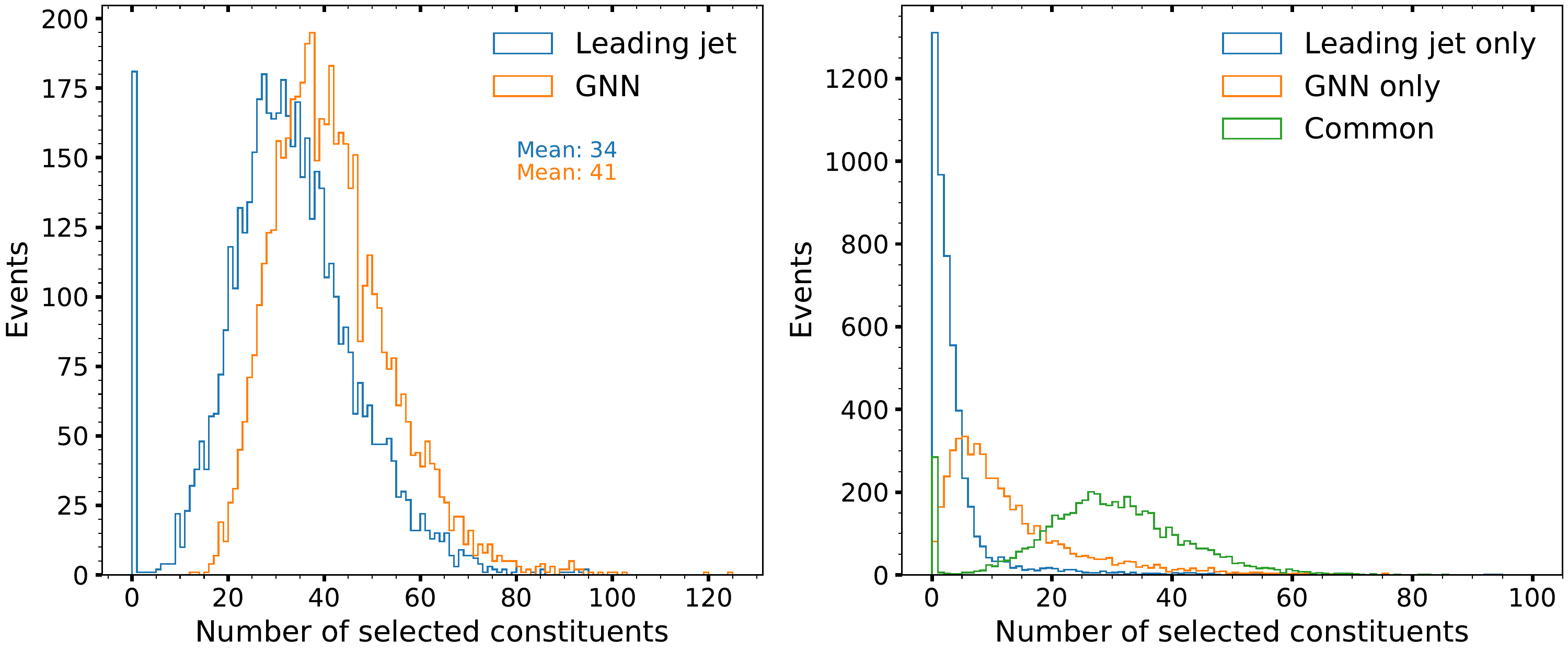}
    \caption{Left: comparison of the number of selected constituents by the edge classifier (GNN) and the constituents inside the leading jet constructed from the anti-$k_t$ algorithm. Right: Decomposition of the selected constituents from the two methods into the ones selected by both methods in green, only by the leading jet in blue and only by the edge classifier in yellow.}
    \label{fig:cmp_particles}
\end{figure}

\begin{figure}[htb]
    \centering
    \includegraphics[width=\linewidth]{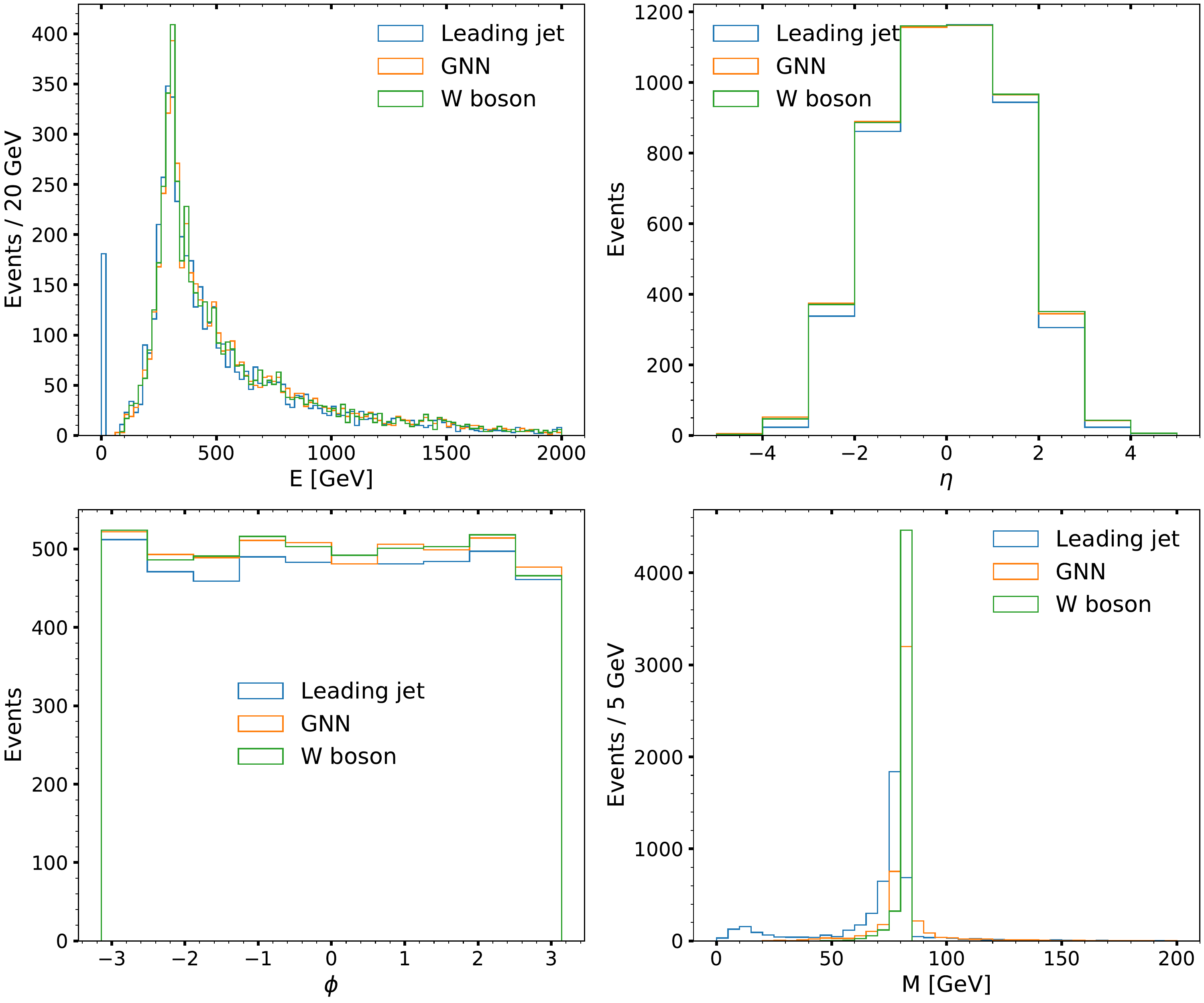}
    \caption{Comparisons of the four-momenta of the reconstructed $W$ boson candidates 
    among the anti-$k_t$ jet clustering in blue, the GNN-based jet clustering in yellow and
    the truth-level $W$ boson in green.  The spike at zero in the top left plot corresponds to events with no anti-$k_t$ jet with $p_T>100$ GeV.  Such events are removed from the other plots.}
    \label{fig:cmp_kinematic}
\end{figure}

\begin{figure}[htb]
    \centering
    \includegraphics[width=0.49\linewidth]{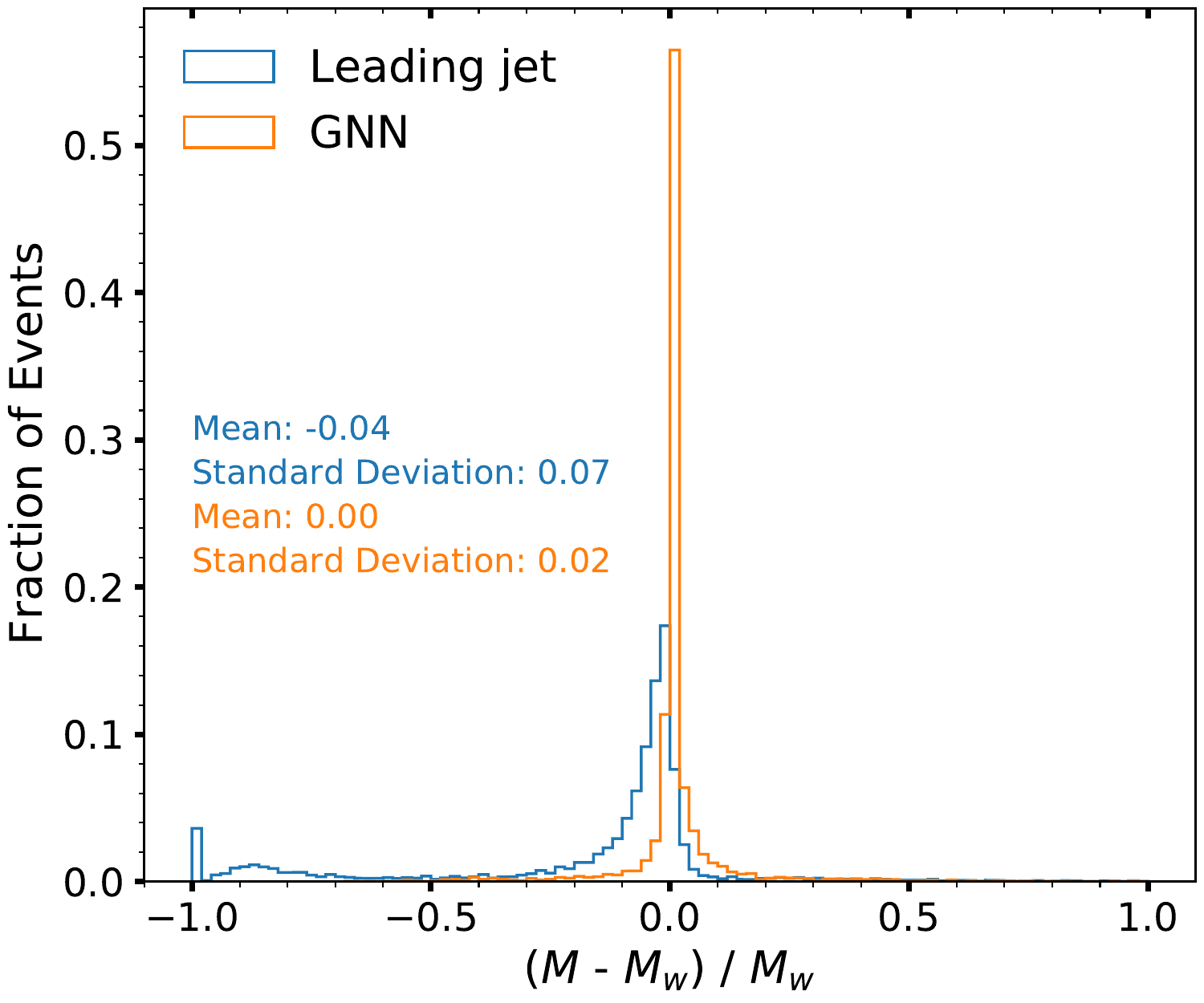}
    \includegraphics[width=0.49\linewidth]{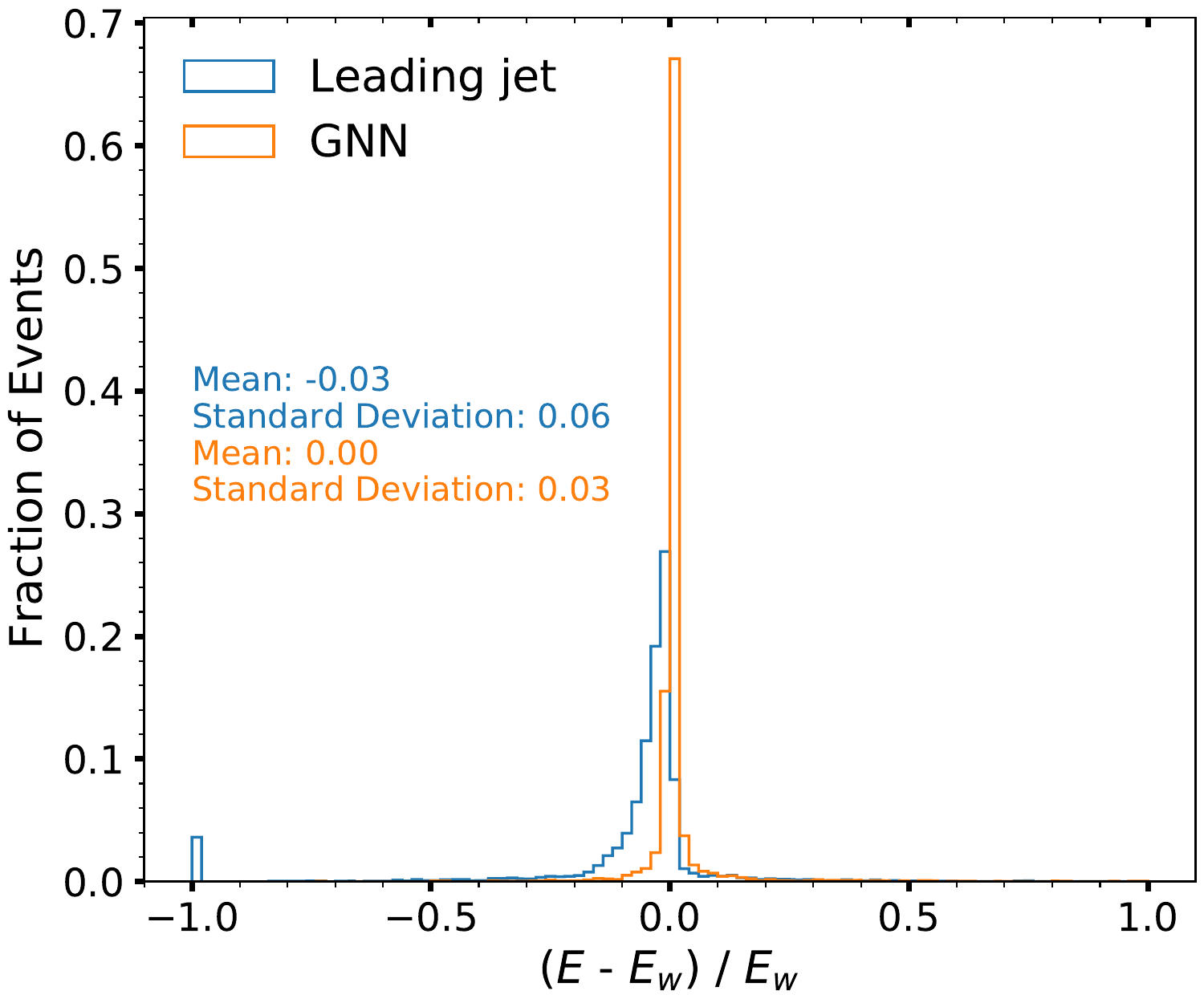}
    \caption{Comparison of the mass resolution (left) and the energy fraction (right) between the anti-$k_t$ jet clustering and the GNN-based
    jet clustering. The mean and standard deviation are calculated in the range of -0.25 and 0.25 in all cases.
    }
    \label{fig:efrac_resolution}
\end{figure}

The nodes that are connected by the edges passing a threshold of 0.5 are considered as the hadrons coming from the $W$ bosons. The four-momenta of the reconstructed $W$ boson is the sum of these surviving hadrons. Figure~\ref{fig:cmp_particles} compares the number of hadrons selected by the edge classifier and the anti-$k_t$ algorithm. On average, the number of hadrons selected by the edge classifier is about 20\% more than that the anti-$k_t$ jet includes, disregarding the events with no anti-$k_t$ jet with $p_T > 100$~GeV. There are also many particles chosen by one algorithm but not by the other.  It will be interesting in the future to examine the properties of such particles to identify which features the GNN is learning differently than anti-$k_t$ (and vice versa). Furthermore, Figure~\ref{fig:cmp_kinematic} compares the kinematic distributions of the $W$ boson candidates reconstructed from hadrons selected by the edge classifier or the anti-$k_t$ algorithm or the truth-labeled. About 3\% of the time, there is no reconstructed jet with $p_T>100$ GeV, which results in the spike at zero. In addition, the fraction of the reconstructed $W$ energy/mass over the total $W$ energy/mass are compared between the two methods in Figure~\ref{fig:efrac_resolution}. In both cases, the GNN-based method significantly outperforms the anti-$k_t$ based method in reconstructing the boosted $W$ bosons\footnote{There is no correct answer for generic quark jets, but the GNN-based jet clustering is applied to the $q^*$ events and the four-momenta of the reconstructed jet is compared with the leading jet from the anti-$k_T$ algorithm in the Appendix (Fig.~\ref{fig:qcd_kinematics}). There is a small tendency of the jet mass to be near the $W$ mass, but it is not as sharp as for $W$ events.}

\begin{figure}[htb]
    \centering
    \includegraphics[width=0.6\linewidth]{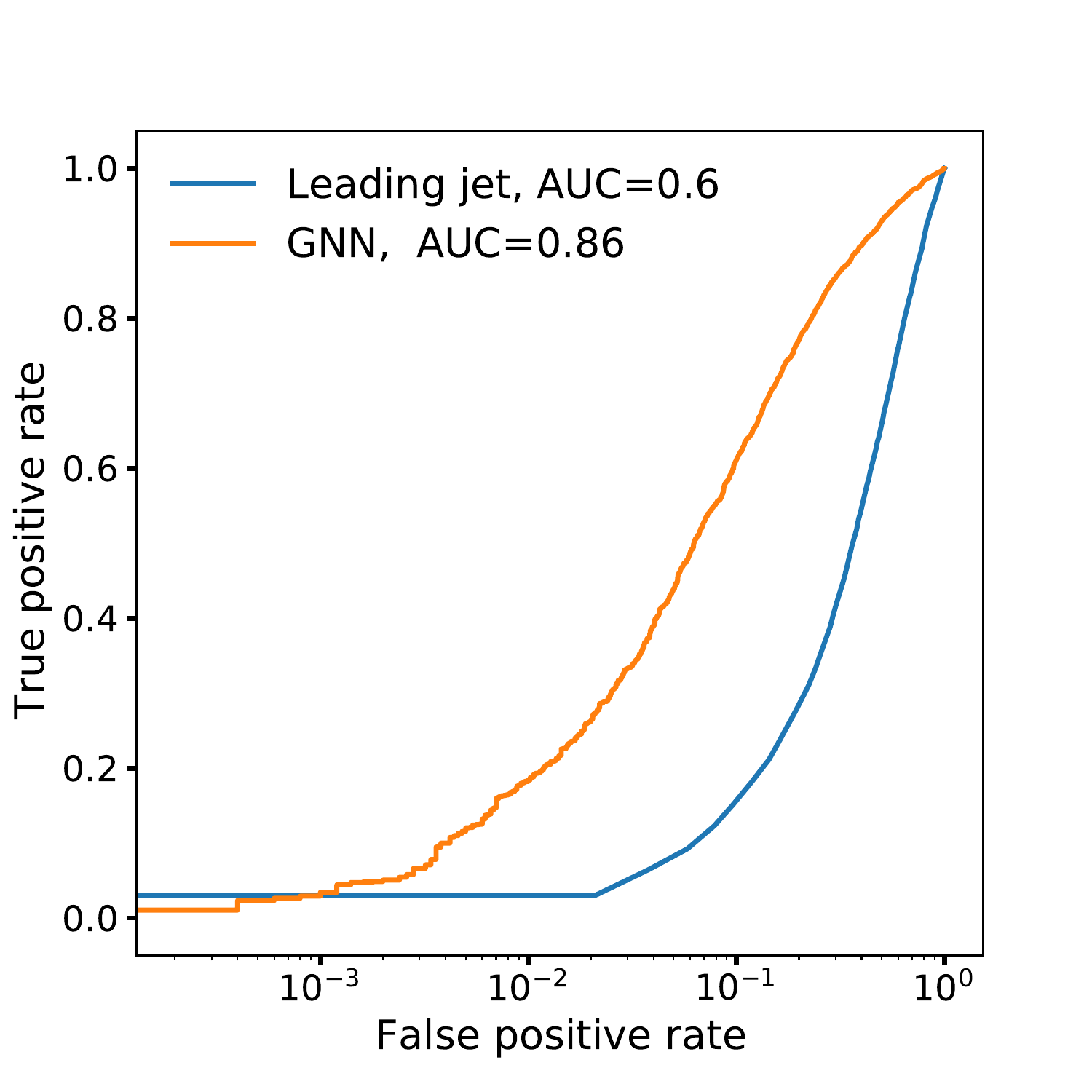}
    \caption{Comparison of the ROC curve from the GNNs trained with the inputs from the anti-$k_t$ based 
    jet clustering and the inputs from the trained edge classifier.  Note that the small inefficiency from the $p_T$ requirement for the anti-$k_t$ jets is not included.}
    \label{fig:cmp_ljet_gnn}
\end{figure}

The event classifiers were trained for 25 epochs for the tGNN and 15 epochs for the eGNN. In both cases, no improvement were seen after these epochs when the GNNs were evaluated on the testing data. Figure~\ref{fig:cmp_ljet_gnn} shows a comparison of the receiver operating characteristic curve (ROC curve) of the two trained GNNs as well as the area under the ROC curve (AUC). The GNN trained with the inputs from the edge classifier outperforms the GNN trained with inputs from the traditional anti-$k_t$ algorithm by more than 40\% in AUC.

\clearpage
\section{Conclusions} \label{sec:conclusions}

Traditional jet clustering based on unsupervised learning has proven to be an effective tool for studying hadronic final states at the LHC.  In particular, the widely-used anti-$k_t$ algorithm is both theoretically and experimentally powerful for studying the SM and searching for physics beyond the SM.  A wide variety of jet substructure techniques using these jets with and without machine learning are being developed and many have already been deployed in data analysis.  However, there is a unique opportunity with color singlet decays to re-examine the construction of jets.

In particular, we have exploited the precise mapping between color singlet particles and final-state hadrons to constructed a supervised jet clustering based on graph neural network.  These jets match the kinematic properties of true $W$ bosons more precisely than the leading anti-$k_t$ jet.  Furthermore, we have shown that there is more information contained in the graph network jets about the originating particle than anti-$k_t$ jets.  In particular, a classifier trained using jet constituents to distinguish $W$ boson jets from quark jets is more effective for GNN jets than for anti-$k_t$ jets.

This work marks the beginning of a new exploration in jet physics to use machine learning to optimize the construction of jets and not only the observables computed from jet constituents.  Tagging Lorentz-boosted color singlet jets is an integral part of measurement and search efforts at the LHC and so further developments in this area have a significant potential to enhance the sensitivity of the LHC physics program.  A variety of further studies will be required to integrate supervised jets into the experimental workflow.  In particular, future work will investigate how event topology effects GNN jets (i.e. what happens when there are more ($W$) jets in the event).  Furthermore, it is important to study the impact of detector-effects and to investigate how well such jets could be calibrated, including pileup stability.


The studies presented in this paper have only considered boosted $W$ bosons, but the same ideas could be applied to any color-singlet particles and it will be interesting to see how GNN jets can be integrated with additional information such as $b$-jet tagging in the case of Higgs bosons.  Examining the structure of the supervised jets may also provide useful physical insight about where the information about the initiating particle is embedded in the event radiation pattern.  Finally, it may be that the ultimate performance is achievable when supervised learning is combined with unsupervised techniques and this could lead to new insight for traditional quark and gluon jet reconstruction.


\section*{\label{sec::acknowledgments}Acknowledgments}


We would like to thank Andrew Larkoski, Zach Marshall, Ian Moult, and Jesse Thaler for useful feedback on the manuscript.

This research used resources of the National Energy Research Scientific Computing Center (NERSC), a U.S. Department of Energy Office of Science User Facility operated under Contract No. DE-AC02-05CH11231.

\bibliography{refs,HEPML,graph}
\bibliographystyle{utphys}

\clearpage

\appendix

\begin{figure}
    \centering
    \includegraphics[width=\linewidth]{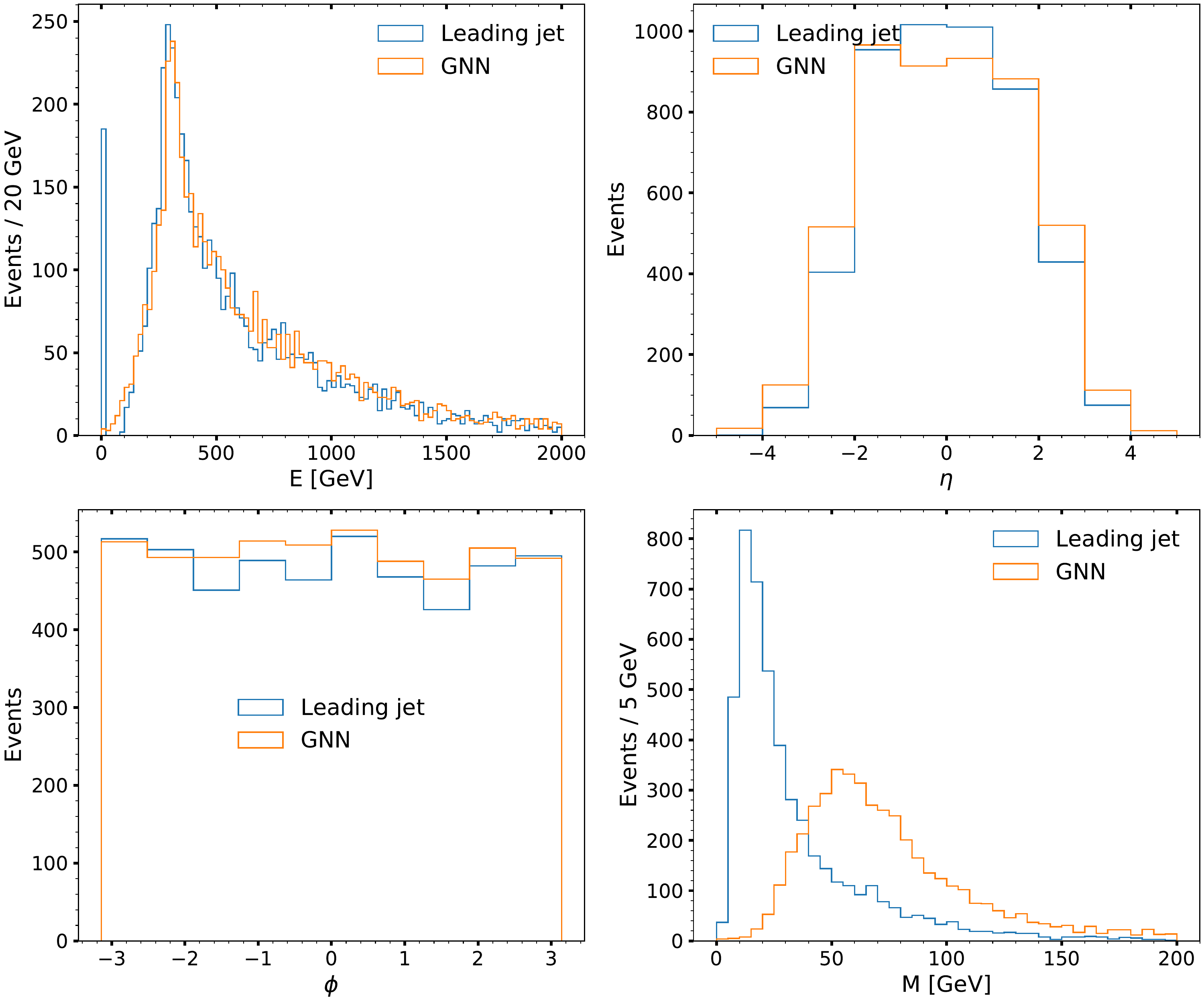}
    \caption{Comparisons of the four-momenta of the reconstructed jet for the $q^*$ events between the anti-$k_T$ jet clustering and the GNN jet clustering.}
    \label{fig:qcd_kinematics}
\end{figure}

\end{document}